\documentclass[12pt,preprint]{aastex}
\usepackage{graphicx,subfigure,amsmath}
\providecommand{\e}[1]{\ensuremath{\times 10^{#1}}}
\newcommand\gsim{\,\lower3pt\hbox{$\sim$}\llap{\raise2pt\hbox{$>$}}\,}
\newcommand\lsim{\,\lower3pt\hbox{$\sim$}\llap{\raise2pt\hbox{$<$}}\,}
\providecommand{\apj}{ApJ}
\providecommand{\apjl}{ApJL}
\providecommand{\nat}{Nature}
\providecommand{\sci}{Science}  
\providecommand{\solphys}{Sol. Phys.}

\begin{document}
\title{A Theory on the Convective Origins of Active Longitudes on Solar-like Stars}

\author{Maria A. Weber\altaffilmark{1}$^{,}$\altaffilmark{2}, Yuhong Fan\altaffilmark{1}, and Mark S. Miesch\altaffilmark{1}}
\affil{High Altitude Observatory, National Center for Atmospheric
Research\altaffilmark{1}, 3080 Center Green Drive, Boulder, CO 80301}

\altaffiltext{1} {The National Center for Atmospheric Research
is sponsored by the National Science Foundation}
\altaffiltext{2}{Graduate Student in the Department of Physics, Colorado State University, Fort Collins, CO 80523, USA}

\begin{abstract}
Using a thin flux tube model in a rotating spherical shell of turbulent, solar-like convective flows, we find that the distribution of emerging flux tubes in our simulation is inhomogeneous in longitude, with properties similar to those of active longitudes on the Sun and other solar-like stars.  The large-scale pattern of flux emergence our simulations produce exhibits preferred longitudinal modes of low order, drift with respect to a fixed reference system, and alignment across the Equator at low latitudes between $\pm15^{\circ}$.  We suggest that these active-longitude-like emergence patterns are the result of columnar, rotationally aligned giant cells present in our convection simulation at low latitudes.  If giant convecting cells exist in the bulk of the solar convection zone, this phenomenon, along with differential rotation, could in part provide an explanation for the behavior of active longitudes.       
\end{abstract}


\section{Introduction}

For longer than half a century, it has been observed that solar active regions tend to emerge near the location of previous or currently existent magnetic flux \citep{eigen45,bumba65,gaiz83,casten86,brouwer90,harvey93}.  Solar observations also show that the emergence of active features is distributed inhomogeneously in longitude according to sunspot activity, solar flares, and coronal streamers \citep{jetsu97,berd03,zhang08,zhang11,olem09,li11}.  Periodic signals have been observed in the solar wind and geomagnetic activity, which may also be attributed to an inhomogeneous longitudinal distribution of emerging magnetic flux on the solar surface \citep{mursula96,neu00,love12}.  These preferential longitudes of solar activity are commonly referred to as active longitudes, and have been observed on some cool, active stars and young solar analogs \citep{olah91,jarv05,lanza09,garcia11}.  

The Sun typically has two active longitudes separated by 180$^{\circ}$ \citep{uso05,zhang11}, although there may be upwards of four or more active longitudes per rotation near solar maximum, and even as few as one or none during solar minimum \citep{detoma2000}.  Active longitudes are also fairly long-lived, with lifetimes of up to seven rotations \citep{detoma2000}, while \citet{berd03} suggest that active longitudes can persist longer than a century.  Owing to the observed North/South asymmetry of solar activity cycles  \citep{verma93,temmer06}, the Northern and Southern hemispheres often exhibit different magnetic behavior, although it is not uncommon for both hemispheres to exhibit the same active longitude \citep{ben99,detoma2000}.  Furthermore, these bands of flux emergence migrate with respect to a rigidly rotating frame \citep{uso07}, and appear to propagate prograde near the Equator at a rate faster than the Carrington sidereal rate \citep{ben99}.

The physical mechanisms which give rise to the active longitude phenomenon still remain relatively unknown, although there exist a few theories. \citet{ruz98} suggests that the localization of a non-axisymmetric mean magnetic field at the base of the convection zone can result in clustering of active regions at a particular longitude.  A fluctuating magnetic field super-imposed on the non-axisymmetric mean field can experience an amplification of its field strength at the longitudinal position of the mean field enhancement.  Subsequently, the magnetic flux loop will rise to the surface provided the field strength is large enough for the onset of the magnetic buoyancy instability.  Non-axisymmetric toroidal magnetic fields that produce buoyant loops have recently been found in convective dynamo simulations by \citet{nelson11,nelson13}.  Conversely, \citet{dik05} show that the toroidal mean magnetic field does not have to be enhanced at a particular longitudinal position in order for active regions to appear at preferred longitudes.  Using a shallow-water model of the tachocline, they find that magnetohydrodynamic (MHD) instabilities can simultaneously produce tipping instabilities of the toroidal magnetic field bands and variations in the thickness of the tachocline material.  A correlation between the tipped toroidal band and a bulge of the tachocline material can force magnetic fields in to the less dense layers of the convection zone, where they will continue to rise because they are more buoyant than the surrounding medium.     

\citet{simon68} allude to the fact that large convecting cells $\sim$ 300,000 km in diameter may be responsible for what \citet{bumba65} call 'complexes of activity', which are clusters of active regions compact spatially in latitude and longitude.  These 'giant cells' were posited by \citet{simon68} as an efficient mechanism of heat transport over multiple density scale heights.  Their observational existence \citep[e.g.][]{hath96,beck98} is supported by giant convecting cellular structures present in three-dimensional simulations of turbulent stellar convection, which align with the rotation axis at low latitudes, remain coherent for at least a rotation period or longer, and propagate prograde near the equator \citep{miesch08,bess11}.  However, many other attempts to detect giant cells on the Sun have failed \citep[e.g.][]{labonte81,snodgrass84,chiang87}, as their signature is difficult to extract from those of granulation ($\sim$ 1,000 km in diameter) and super-granulation ($\sim$ 30,000 km in diameter), upon which the giant cell signature may be super-imposed. 

Using a thin flux tube model in a rotating spherical shell of turbulent, solar-like convective flows as described in Section \ref{sec:model}, we find evidence which suggests that giant convecting cells (on the order of $\sim$50 Mm - 100 Mm in diameter) in the bulk of the solar convection zone can organize buoyantly rising flux tubes such that large-scale emergence patterns at or near the solar surface are formed.  We describe how we extract large-scale flux emergence patterns from our simulations in Section \ref{sec:extract}, and properties of the resulting flux emergence patterns are presented in Section \ref{sec:results}. In performing our analysis, we find that the large-scale flux emergence patterns exhibit some similarities to properties of active longitudes on the Sun.  While the actual physical mechanism responsible for the active longitude phenomenon on the Sun and other stars is most likely a complex process involving contributions from multiple sources, we discuss in Section \ref{sec:discuss} how convection alone can organize flux emergence in a large-scale way.


\section{Model Description}
\label{sec:model}
It is believed that magnetic flux emergence at the solar surface is the result of buoyantly rising magnetic flux tubes generated by a dynamo mechanism at or near the base of the convection zone \citep{spiegel80,ball82,mi92,gilman00,char10}.  The thin flux tube model has been used by a number of authors to model how thin, isolated magnetic flux tubes traverse the solar convection zone \citep{spruit81,mi86,cheng92,ferriz93,longcope97,fan09}.  As described in \citet{weber11} (Paper 1) and \citet{weber12} (Paper 2), we employ a thin flux tube model subject to a turbulent, time-dependent, solar-like convective velocity field to more accurately study the rise of active-region-scale flux tubes.  In this Paper, we use the simulations described in Paper 2 to investigate longitudinal patterns of flux tube emergence.  These simulations considered a range of flux tube magnetic field strengths from 15 kG (equipartition) to 100 kG (super-equipartition), at latitudes ranging from $2^{\circ}$ to $40^{\circ}$ around the equator in both hemispheres, with magnetic flux values of $10^{20}$ Mx to $10^{22}$ Mx.  This range of magnetic flux is typical of ephemeral regions and pores to the strongest sunspots \citep{zwaan87}. We perform ten simulation sets (three more have been added in addition to those computed in Paper 2) sampling different time ranges of the convective flow field for each initial latitude, field strength, and magnetic flux we consider, for a total of 8640 flux tubes.  Each set of flux tubes are released at the base of the convection zone at the same starting time, although do not interact with each other (i.e. are isolated), and  are allowed to evolve until some portion of the tube reaches the top of the simulation domain.  The flux tube release times for the ten sets are arbitrary, but are at least separated by the convective turnover time scale of the convection simulation, which is $\sim30$ days.  In this way, the flux tubes are able to sample significantly different portions of the convective velocity field.  

The thin flux tube model we use and the associated equations which describe the evolution of the flux tube have been described in detail in Paper 1 and Paper 2.  The three-dimensional global convection simulation in which the flux tube evolves is computed separately from the thin flux tube simulation using the Anelastic Spherical Harmonic (ASH) code, as described by \citet{mieschetal2006}.  This time-dependent convective velocity field, which is computed relative to the rotating frame of reference with angular velocity $\Omega_{0} = 2.7 \times 10^{-6}$ rad s$^{-1}$ , impacts the thin flux tube through its drag force term.  In the anelastic approximation, the velocity of convective flows is taken to be much slower than the speed of sound in the fluid, and convective flows and thermal variations are treated as a linear perturbation to a background state taken from a one-dimensional solar structure model.  The computed convective velocity field captures giant-cell convection, and associated mean flows such as meridional circulation and differential rotation, in a rotating convective envelope spanning $r = 0.69 R_{\odot}$ to $r = 0.97 R_{\odot}$ (4.8\e{10} cm to 6.75\e{10} cm from Sun center).  For a more detailed description of this particular convection simulation, see Paper 1.  

\begin{figure}
\includegraphics[angle=90,scale=1.]{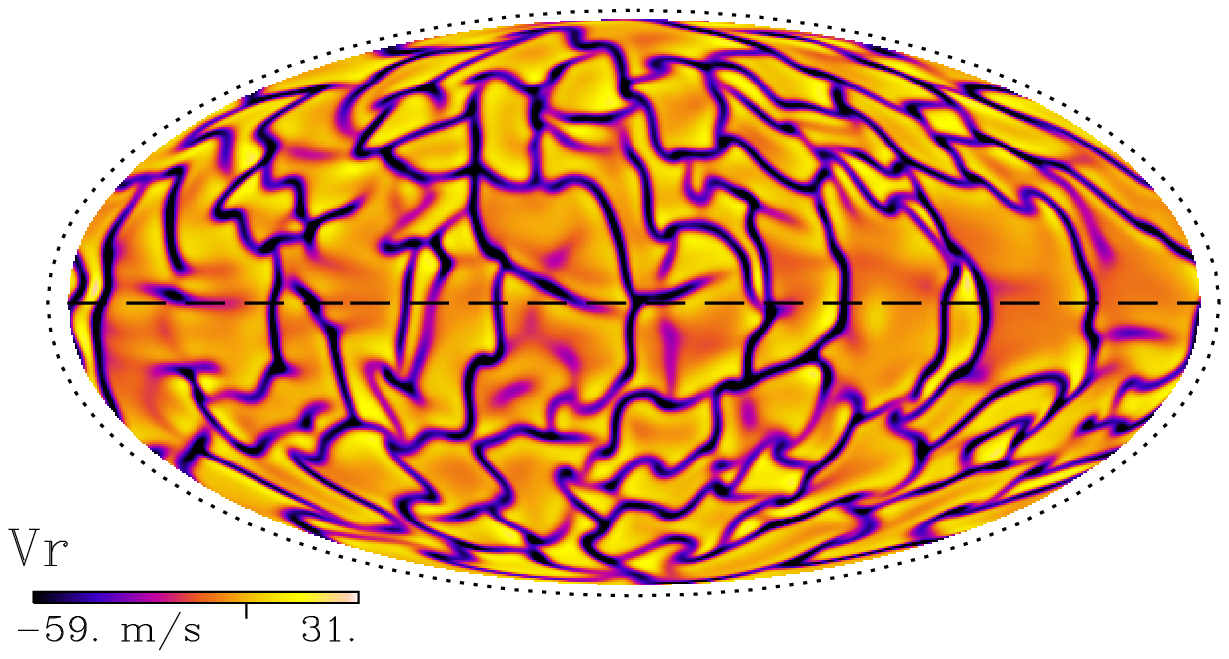}
\caption{Snapshot of convective radial velocity at a depth of 23 Mm below the solar surface (r=0.97R$_{\odot}$) in a Mollweide projection.  This figure shows strong downflow lanes (purple/blue) at the boundary of giant convective cells.  Also known as banana cells, the structures at low latitudes are rotationally aligned and propagate prograde.  The dotted line is the solar radius r=R$_{\odot}$.  The radial velocity approaches zero at the upper boundary of the simulation, so the velocity amplitudes shown in this figure are lower than in the middle convection zone.}
\label{fig:mollweide}
\end{figure}

\begin{figure}
\epsscale{1.}
\plotone{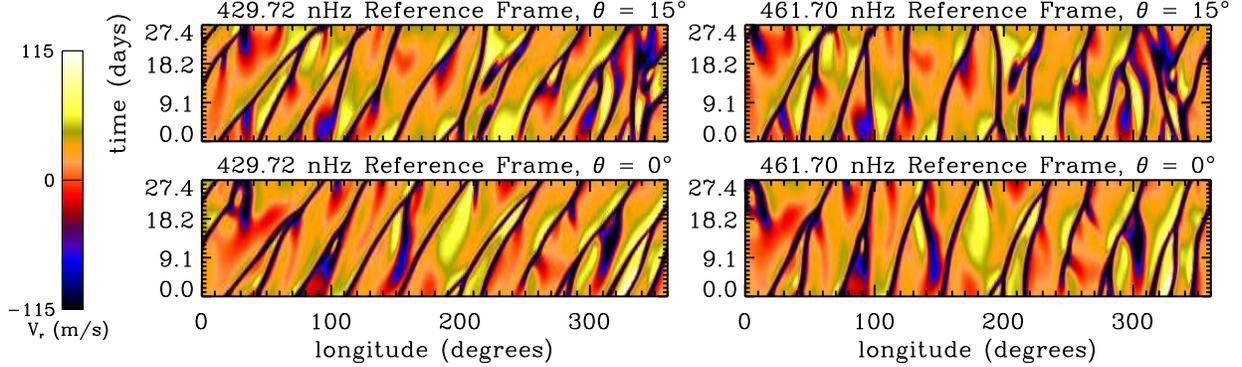}
\caption{Radial velocity at the Equator (bottom) and 15$^{\circ}$ latitude (top) plotted as a function of longitude and time (i.e. ($\phi$,t) diagrams), at a depth of r=$0.95R_{\odot}$ for a 27.4 day period in the reference frame rotating at angular velocity $\Omega_{0}/2\pi$ (429.72 nHz, left), and in the faster ASH equivalent Carrington frame $\Omega_{AC}/2\pi$ (461.70 nHz, right).  In the $\Omega_{0}/2\pi$ reference frame,  the rightward tilt of the dark blue downflow lanes indicates a prograde propagation of the downflow lanes at low latitudes within $\pm15^{\circ}$ latitude of the Equator, although their rate of prograde propagation decreases as the latitude increases.  In the faster rotating reference frame $\Omega_{AC}/2\pi$, downflow lanes still propagate prograde near the Equator, but at a slower relative rate than in the $\Omega_{0}/2\pi$ reference frame. At higher latitudes in the $\Omega_{AC}/2\pi$ reference frame, some downflow lanes no longer appear to propagate prograde, remaining almost stationary or moving retrograde.}
\label{fig:ash_strip}
\end{figure}

\begin{figure}
\includegraphics[angle=90,scale=.5]{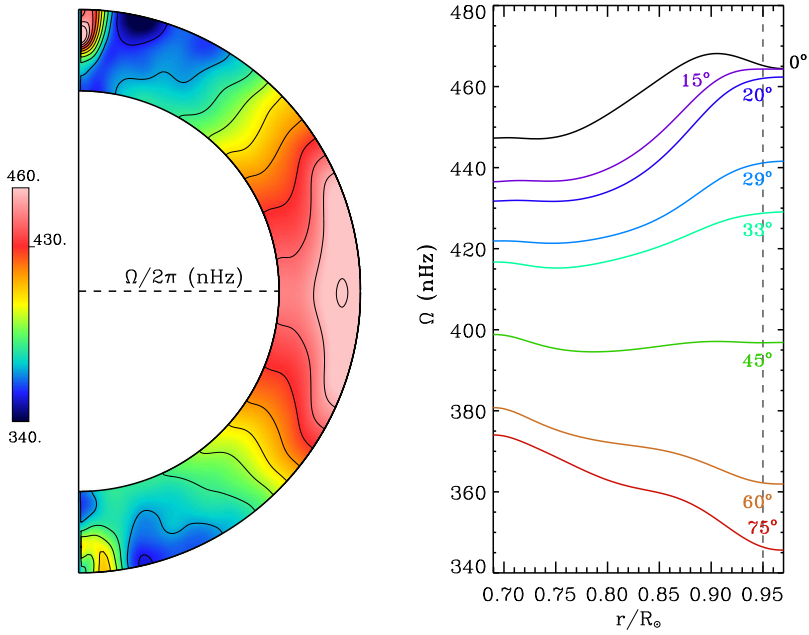}
\caption{(Left) Angular velocity in the convection simulation, averaged over longitude and time (time interval of 755 days).  Color table saturates at the values indicated, with extrema ranging from 326 nHz - 468 nHz.  (Right) Angular velocity of the convection simulation at specific latitudes as a function of radius.  At 0.95R$_{\odot}$, the angular velocity $\Omega/2\pi$ is 461.70 nHz at 20$^{\circ}$ latitude, 440.64 nHz at 29$^{\circ}$, and 429.72 nHz at 33$^{\circ}$ latitude.}
\label{fig:DR}
\end{figure}

A typical giant-cell convection pattern at a depth of 25 Mm below the solar surface is shown in Figure \ref{fig:mollweide}.  Broad upflow cells are surrounded by narrow downflow lanes, which can reach maximum downflow speeds of nearly 600 m s$^{-1}$ at a mid-convection zone depth of $\sim86$ Mm below the surface.  Columnar, elongated downflow lanes associated with these giant cells align preferentially with the rotation axis at low latitudes, and propagate in a prograde direction relative to the polar regions, due in part to differential rotation and an intrinsic phase drift similar to traveling Rossby waves \citep{miesch_toomre2009}.  Such structures also can remain coherent for at least a rotation period or longer.  The prograde propagation and coherency of the giant-cell-associated downflows is exhibited in the ($\phi$,t) diagrams of Figure \ref{fig:ash_strip}.  This Figure shows a strip of radial velocity at the Equator (bottom) and 15$^{\circ}$ (top) for $\sim27$ consecutive days, or about one rotation period.  In Figure \ref{fig:ash_strip}, the ($\phi$,t) diagrams are also shown in two different rotating reference systems, with the two panels on the left shown in the reference frame co-rotating with the mean angular velocity $\Omega_{0}/2\pi=429.72$ nHz of our simulation, and the two panels on the right in the reference frame rotating at a faster rate of $\Omega_{AC}/2\pi=461.70$ nHz.  These features dominate the convective Reynolds stress, aiding in the maintenance of a strong differential rotation. The total angular velocity $\Omega/2 \pi$ (with respect to the inertial frame) is solar-like, and decreases monotonically from $\approx$470 nHz at the equator to $\approx$330 nHz at the poles, and exhibits nearly conical contours at mid-latitudes (see Figure \ref{fig:DR}), as observed in the solar convection zone via helioseismic inversions \citep{thompson2003}.     

\begin{figure}
\includegraphics[scale=.9]{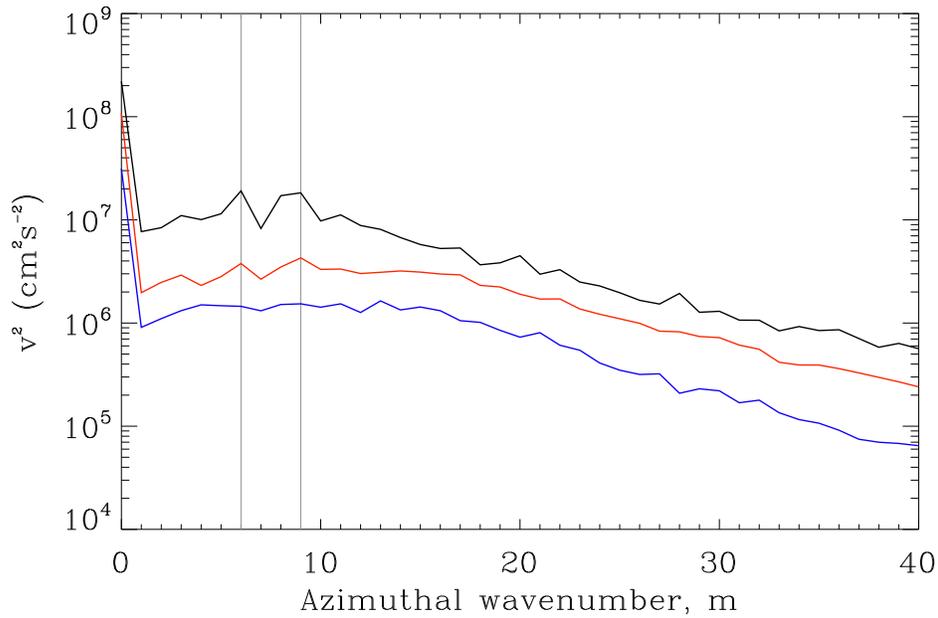} \\
\caption{Spectral decomposition of the velocity variance of the ASH convection simulation in terms of azimuthal (longitudinal) wavenumbers, computed at the Equator for three shells of radii: 0.95R$_{\odot}$ (black), 0.83R$_{\odot}$ (red), and 0.73R$_{\odot}$ (blue).  All curves are averaged over the Northern and Southern hemisphere for a time interval of 19 rotation periods (26.93 days each).  Prominent peaks in the spectra at wavenumbers of $m=6$ and $m=9$, corresponding to the gray vertical lines, are representative of the periodic banana-like convective cells present in the simulation.}
\label{fig:wavenumbers}
\end{figure}

A spectral decomposition of the velocity variance in terms of azimuthal (longitudinal) wavenumbers demonstrates the distribution of contributing convective modes in the ASH simulation.  This is shown in Figure \ref{fig:wavenumbers} for three radial levels in the convection zone, only at the Equator.  Peaks in the spectra at $m=0$ for the curves in Figure \ref{fig:wavenumbers} are primarily caused by differential rotation, with a small contribution from the meridional circulation.  Other prominent peaks in the spectra in the upper convection zone at r=0.95R$_{\odot}$ are found at azimuthal wavenumbers of 6, 8 and 9.  This reflects the nature of the elongated, periodic, banana-like downflow structures obvious in Figure \ref{fig:mollweide}.  Signatures of these banana cells are also evident near the Equator in the mid-convection zone at r=0.83R$_{\odot}$.  

While this ASH simulation, with a mid-convection zone Raleigh number of 5\e6 and Reynolds number of $\sim50$, is more laminar than some others \citep[e.g.][]{miesch08,jouve09}, it still possesses all of the relevant features necessary to explore the fundamental interactions between thin flux tubes and the mean flows associated with global convection. These features include: asymmetric, rotationally aligned cells at low latitudes (density-stratified banana cells), rapidly-evolving downflows in the upper convection zone at high latitudes dominated by helical plumes, and a strong, solar-like differential rotation.  We believe that large-scale, columnar banana cells must persist even in highly turbulent parameter regimes in order to provide the requisite Reynolds stresses to account for the solar differential rotation.  Therefore, we would not expect the essential results to change significantly with more turbulent convection.

The coupling of the thin flux tube model with the ASH convection simulation results in a large-scale longitudinal emergence pattern of the flux tubes at low latitudes, to which we devote the rest of this paper.


\section{Extracting Large-Scale Flux Emergence Patterns from Simulations}
\label{sec:extract}

\begin{figure}
\epsscale{1}
\plotone{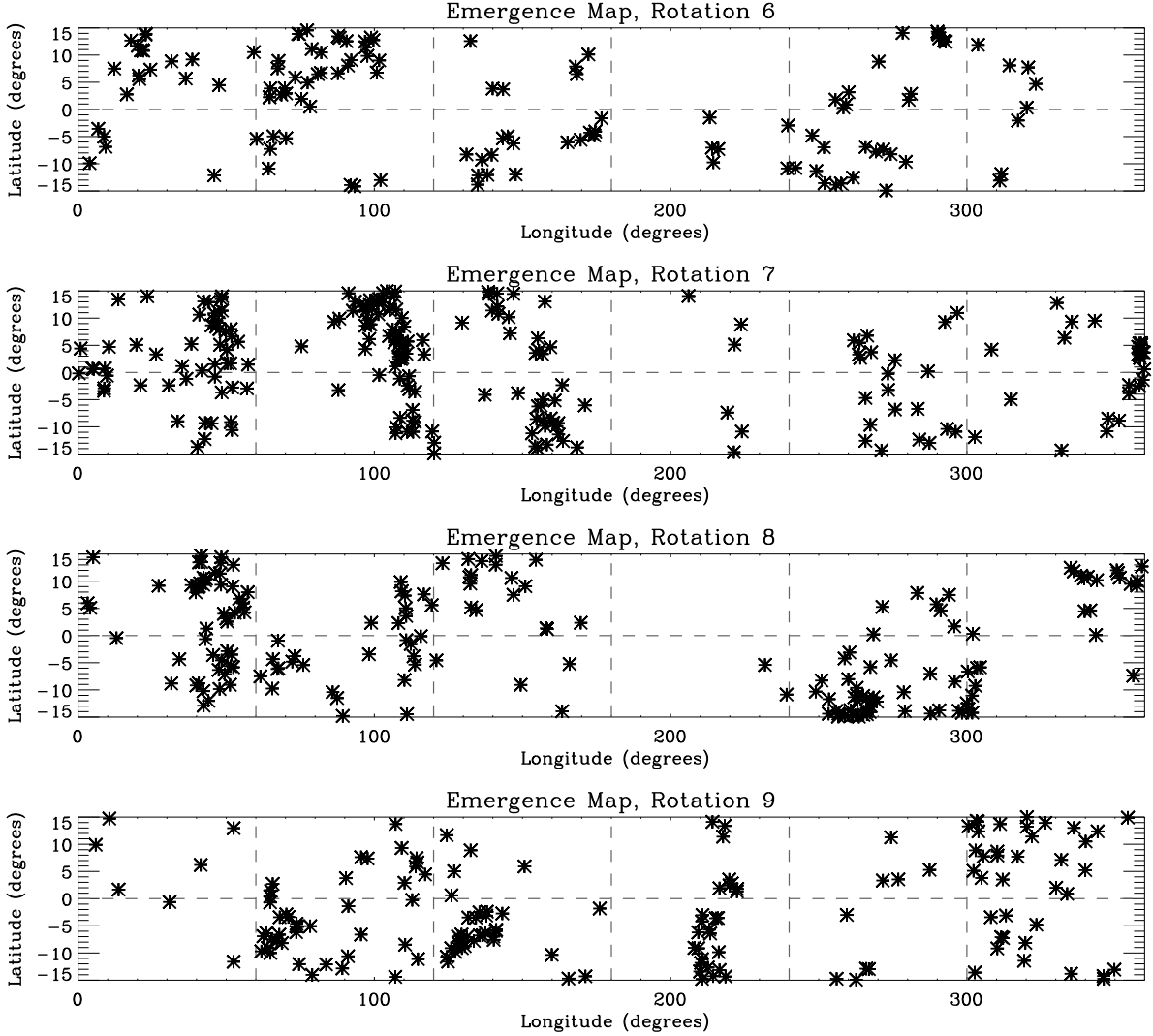}
\caption{Emergence maps for 4 consecutive rotation periods in the reference frame rotating at angular velocity $\Omega_{0}/2\pi$.  The latitudinal ($\theta$) and longitudinal ($\phi$) position of the flux tube apex is plotted for all flux tubes which reach the top of our simulation domain within one rotation period of each other, $\pm15^{\circ}$ from the Equator.  This figure indicates that our simulation produces flux tubes which emerge neither randomly nor uniformly.}
\label{fig:emergence_map}
\end{figure}

Convective downflows and the growth of the magnetic buoyancy instability anchor portions of the flux tube in the overshoot region, allowing buoyant loops to rise through the simulation domain (see Paper 1).  When these simulated flux tubes emerge near the surface, we find that they do not emerge randomly during a rotation period, but rather in distinct longitudinal bands or clusters as depicted in Figure \ref{fig:emergence_map}.  For this Paper, we choose only to investigate flux tube emergence patterns $\pm$15$^{\circ}$ around the equator, so as to focus on the low latitude behavior of these longitudinal bands of flux emergence.  In Figure \ref{fig:emergence_map}, we note that there are wide longitudinal regions void of any flux emergence, and particular longitudinal spans where flux emergence prefers to cluster.  

\begin{figure}
\epsscale{.8}
\plotone{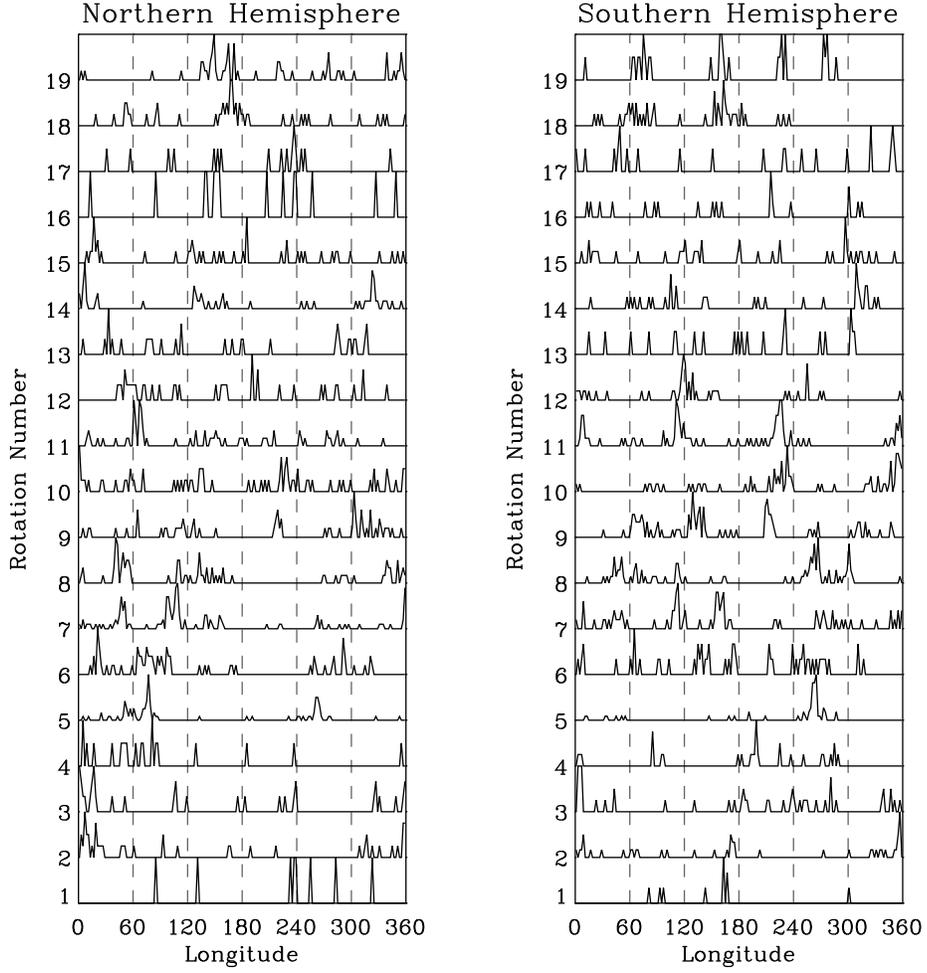}
\caption{Normalized histograms of flux tube apex longitudinal position (emergence histograms) once some portion of the flux tube has reach the top of our simulation domain.  For one rotation period of our simulation, we count the number of flux tubes which emerge within 1 of 180 evenly distributed bins in longitude.  In this Figure, longitudinal coordinates are with respect to a reference frame rotating at the angular velocity $\Omega_{0}/2\pi$.  This is done for the Northern (left) and Southern (right) hemispheres for flux tubes that emerge within $\pm15^{\circ}$ of the equator for 19 consecutive rotations.}
\label{fig:histo}
\end{figure}

To more quantitatively investigate the longitudinal flux emergence distribution pattern our simulation produces, we create 'emergence histograms' wherein we count the number of flux tubes that emerge within one of 180 evenly distributed longitudinal bins during a particular rotation period.  We only count the first portion of each tube that reaches the simulation upper boundary, therefore we count each flux tube only once.  This is done separately for the Northern and Southern hemispheres for each of the 19 consecutive rotation periods considered for this study, only for flux emergence within $\pm15^{\circ}$ of the Equator.  These histograms are shown in Figure \ref{fig:histo} for the reference frame rotating at angular velocity $\Omega_{0}/2\pi$.  The choice of 180 bins of each 2$^{\circ}$ in longitude is a rather arbitrary one.  We have chosen to use small longitudinal bins in order to more accurately capture the drift rate of the flux emergence patters as discussed in Section \ref{sec:prograde}.       

In Figures \ref{fig:emergence_map} and \ref{fig:histo}, we have used $\Omega_{0}/2\pi=429.72$ nHz as the angular velocity of the rotating reference frame, with a rotation period of 26.93 days.  This corresponds to the angular velocity of fluid flows in the ASH simulation at latitude $\theta=33^{\circ}$ and radius r=0.95R$_{\odot}$.  However, when investigating active longitude behavior on the Sun, the sidereal Carrington rotation rate of 456.03 nHz (25.38 days) is often used as the rotating frame of reference, because active longitudes drift relative to this reference frame.  This rotation rate corresponds to the rotation of sunspots at $\sim20^{\circ}$ latitude on the solar surface \citep{thompson2003}.  The observed rotation rate of sunspots at all emergence latitudes also closely follows the plasma rotation rate of the Sun at a depth of r=0.95R$_{\odot}$ for the same latitude (see Fig. 9 in Paper 2).  Therefore, we identify what we call the 'ASH equivalent Carrington rate' as the rotation rate of the convection simulation at latitude $\theta=20^{\circ}$ and radius r=0.95R$_{\odot}$, which is $\Omega_{AC}/2\pi=461.72$ nHz (25.07 days).  We also identify a third reference frame that rotates at the average drift rate of the flux emergence patterns, which we identify and discuss in section \ref{sec:prograde}.  (See the right side of Figure \ref{fig:DR} for the angular velocity profile of the convection simulation at specified latitudes.)  In order to generate emergence histograms in the two reference frames which do not rotate at angular velocity $\Omega_{0}/2\pi$, we translate the flux emergence longitude coordinates in to the new reference frame, using the new rotation periods to identify 19 consecutive rotations that all start at the same reference time.  We utilize these histograms in Section \ref{sec:results} to characterize the large-scale pattern of flux emergence generated by our simulations.     

Our study is limited to only 19 rotation periods due to the duration of the three-dimensional volume cube data set of our convection simulation, as well as the rise times of our flux tubes.  It is important to note that the methods used in this Paper do not produce a solar cycle dynamo simulation.  This model allows us to investigate how convection can alter the rise of many magnetic flux tubes with various initial conditions.  As such, this simulation will not allow us to investigate changes which occur in flux emergence during the course of a solar cycle.  Due to this, our simulation is not capable of producing the solar butterfly diagram, and we will often have more flux tubes which emerge at the top of our simulation domain than may actually emerge on the solar surface during a rotation period.  

Throughout the Results section, we adopt a 99.7$\%$ confidence level to indicate the significance of our results.  When our analysis reveals results above this value, we are 99.7$\%$ confident the results of our simulation are not due to a random (non-uniform) distribution of flux tube emergence.  For each rotation period and hemisphere, a certain number of flux tubes N emerge within 15$^{\circ}$ of the Equator.  We then create a corresponding array of N elements, representing flux tubes which have random longitudinal emergence positions from 0$^{\circ}$ to 360$^{\circ}$.  These arrays representing random longitudinal flux tube emergence are then subjected to the same treatment throughout the Results section as the longitudinal emergence positions for flux tubes allowed to evolve in convection.  Rather than directly plotting the results from the random emergence position arrays, we compute the standard deviation $\sigma$ of the result as  well as the average $X$, then plot $X+3\sigma$ as our 99.7$\%$ confidence level.       


\section{Active-longitude-like Behavior of Flux Emergence Pattern}
\label{sec:results}

\subsection{Longitudinal Inhomogeneity of Emerging Flux Loops}
\label{sec:long_inhomo}

A distinct longitudinal pattern of magnetic activity is observed on the Sun not only in terms of sunspot activity \citep{detoma2000,berd03}, but also in solar x-ray flares \citep{zhang11} and coronal streamers \citep{li11}.  Interestingly, this phenomena is not unique to the Sun, and has been observed on other young solar analogues such as AB Dor \citep{jarv05} and LQ Hydrae \citep{berd02}.  To capture the longitudinal inhomogeneity of flux tube emergence in our simulations, we calculate the variability coefficient $V$ for each rotation period and hemisphere based on \citet{olem09}.  We begin by identifying the quantity $S_{i}$, which is the longitudinal distribution of relative flux tube emergence over each of nine evenly spaced bins in longitude per a particular rotation period:

\begin{equation}
S_{i}=\frac{9n_{i}}{N},
\end{equation}   

where $n_{i}$ is the number of events in a certain longitudinal interval, and $N$ is the total number of events.  Values of $S_{i}=1$ for each of the nine bins per rotation period implies a homogeneous distribution of flux tube emergence.  We consider only the longitudinal position of the flux tube apex once it has reached the top of our simulation domain.  Next the variability coefficient is computed, which is the sum of absolute deviations of $S_{i}$ from the average value of unity:

\begin{equation}
V=\sum_{i=1}^9 |S_{i}-1|.
\end{equation}

\begin{figure}
\epsscale{1}
\plotone{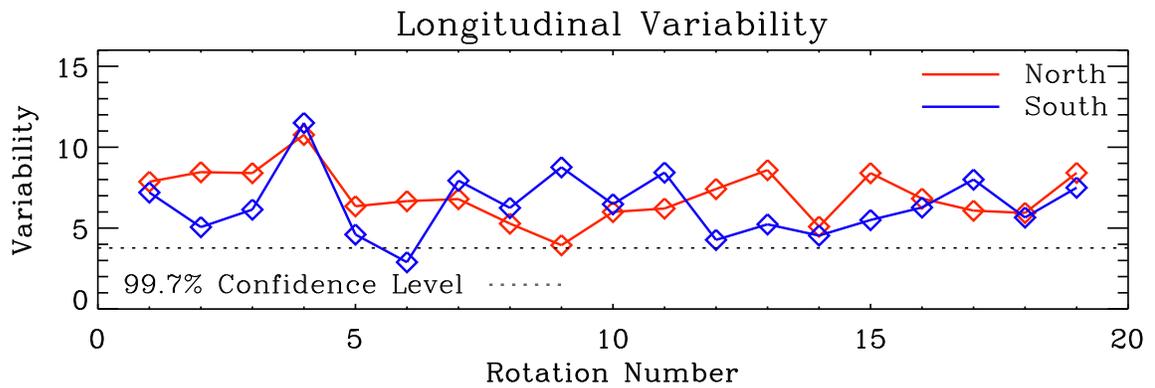}
\caption{Longitudinal variability per rotation period for each hemisphere in reference frame rotating at the angular velocity $\Omega_{0}/2\pi$ of the simulation.  The dotted line is the level above which we are 99.7$\%$ sure the result is not due to a random longitudinal distribution of flux tube emergence.  For the same rotation period, the longitudinal variability can vary significantly for each hemisphere.  }
\label{fig:var}
\end{figure}

The calculated variability coefficient ranges from zero to 16, with the minimum value typical of a homogenous distribution of events in the longitudinal intervals, and the maximum value typical of all events falling in one longitudinal interval.  The longitudinal variability in the Northern and Southern hemispheres for each of 19 rotation periods in the reference frame which rotates at the angular velocity $\Omega_{0}/2\pi$ is shown in Figure \ref{fig:var}.  The dotted line represents the 99.7$\%$ (3$\sigma$) confidence level, above which we are 99.7$\%$ positive the variability is not the result of a longitudinally random (non-uniform) distribution of flux tube emergence.  All of the variability coefficients lie above this line, with the exception of rotation 6 for the Southern hemisphere.  As expected, most rotation periods exhibit a variability coefficient above the 99.7$\%$ confidence level regardless of the choice of reference frame. Therefore, we can reasonably say that the distribution of flux tube emergence in our simulation is not uniform or random in longitude for the majority of the rotation periods we consider.

\begin{figure}
\epsscale{1}
\plotone{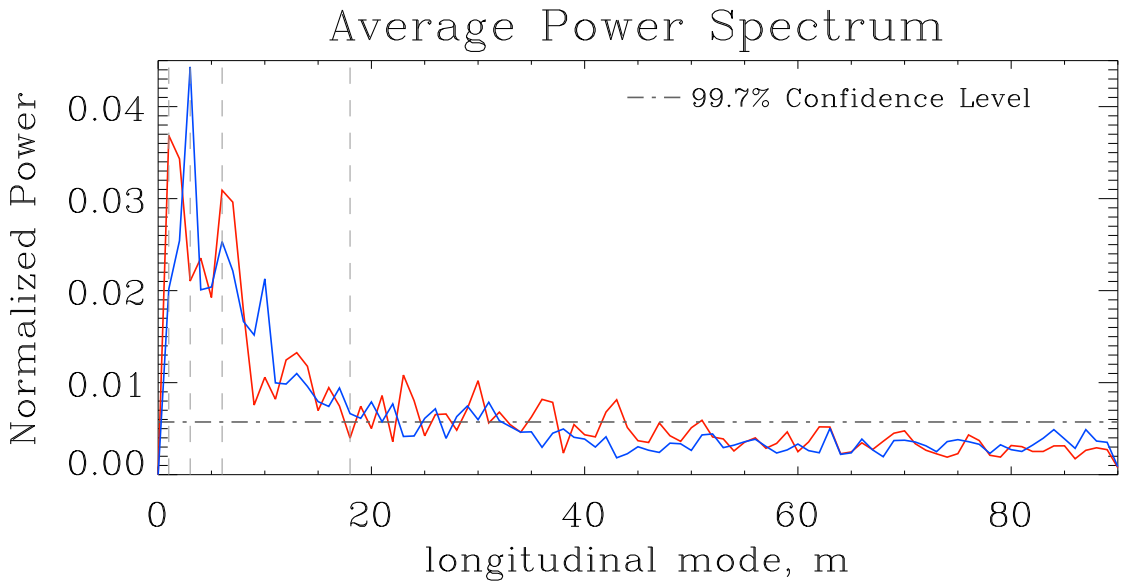}
\caption{Average power spectrum for the Northern (red) and Southern (blue) hemispheres for 19 rotation periods in the reference frame rotating at $\Omega_{0}/2\pi$, considering all magnetic field strengths and magnetic flux.  The power spectrum peaks for a longitudinal mode of $m=1$ in the Northern hemisphere, and $m=3$ in the Southern hemisphere, shown by dashed lines.  In addition, the $m=6$ and $m=18$ modes are also shown.  The dash-dotted line is the level above which we are 99.7$\%$ certain the result is not due to a random distribution of emerging flux tubes in longitude.  The nature of our convection simulation results in flux emergence at preferred longitudinal modes.}
\label{fig:power}
\end{figure}

The above analysis shows that more flux tubes may emerge per a certain longitudinal span for each hemisphere, i.e. there is a significant non-random clustering of flux tube emergence.  On the Sun, there are typically two active longitudes separated by 180$^{\circ}$ \citep{uso05,zhang11}, although this number may increase to as many as four or more near solar maximum \citep{detoma2000}.  In order to identify whether or not our simulation produces flux tubes which emerge with a preferred longitudinal mode, we perform a power spectrum analysis for each of the emergence histograms in Figure \ref{fig:histo}.  We then take the average of these power spectra for both the Northern and Southern hemispheres separately in order to identify an overall trend for the 19 rotation periods we consider, which is shown in Figure \ref{fig:power} in the reference frame rotating at $\Omega_{0}/2\pi$.  

While we identify relative maxima of the power spectra in the $\Omega_{0}/2\pi$ reference frame at low spatial frequencies, corresponding to a low order number of longitudinal modes, we need to compare these results to a random distribution in longitude of flux tube emergence to assess the significance of these maxima.  The dash-dotted line in Figure \ref{fig:power} represents the level above which we are 99.7$\%$ positive the result of the averaged power spectrum is not due to a longitudinally random distribution of emerging flux tubes.  The averaged power spectra for both hemispheres is above the 99.7$\%$ confidence level for spatial frequencies of 0.05 deg$^{-1}$  and less.   This spatial frequency corresponds to a longitudinal mode of $m=18$, indicating that lower order modes ($m\le18$) are the significant modes present in the flux tube emergence patterns.  

Dominant peaks in the averaged power spectra occur at longitudinal modes of $m=1$ for the Northern hemisphere, and $m=3$ for the Southern hemisphere.  This corresponds to one active longitude in the Northern hemisphere, and three in the Southern hemisphere on average for a particular rotation period.  The different dominant modes in the power spectra for both hemispheres reflect the fact that the convection pattern is not perfectly symmetric across the Equator on a short time scale.  Additionally, the power spectra of the convection simulation does not show significant power in any mode below $m=6$ (see Fig. \ref{fig:wavenumbers}).  It is likely that there are substantial differences between the Northern and Southern hemisphere convection at these modes. However, both the Northern and Southern hemispheres do have corresponding relative maxima at $m=6$.  This is most likely related to the $m=6$ peak in the power spectra of the convection simulation shown in Fig. \ref{fig:wavenumbers} at r=0.95R$_{\odot}$, indicating that in the upper convection zone, on average six strong, periodic downflow lanes exist.  

Considering the conclusions of Paper 1 and Paper 2, we suggest that the most likely candidate magnetic field strengths for solar dynamo generated flux tubes is on the order of $\gtrsim40$ kG, but most likely not exceeding $\sim100$ kG.  With this in mind, if we perform the power spectrum analysis considering only 40 kG - 100 kG flux tubes, peaks in the power spectra remain at the same longitudinal mode as the case where we consider all magnetic field strengths together.  Maximum peaks in the power spectra also remain the same regardless of the choice of reference frame. 

In the absence of convection, flux tubes will emerge randomly distributed in longitude if they are perturbed with random undular motions to initiate a buoyancy instability.  Therefore, these flux tubes will not rise to create flux emergence patterns with preferred longitudinal modes.  The fact that the flux emergence patterns our simulation produces exhibits low order longitudinal modes suggests that the convective velocity field aides flux tubes in emerging in preferred longitudinal spans, specifically a result of the discrete azimuthal modes of convection, corresponding to strong downflow lanes.  Additionally, both hemispheres do not exhibit the same dominant mode in the power spectrum, indicating that the large-scale nature of the convection simulation is not identical in both hemispheres, nor would we expect it to be on the Sun.

For each rising flux tube, only the longitude of the first rising loop that reaches the simulation upper boundary is included in the analysis.  It is perhaps possible that the 180$^{\circ}$ separation of active longitudes on the Sun is in part a result of a non-axisymmetric $m=2$ mode which a flux tube of large magnetic field strength ($\sim$60 kG -100 kG) may develop due to the non-linear growth of the magnetic buoyancy instability (see Paper 1), which we do not investigate here.  In any case, our simulation suggests that convective flows have some part to play in the preferred modes of active longitude organization.

\subsection{Propagation of Flux Emergence Patterns}
\label{sec:prograde}

While the exact rotation velocity of active longitudes on the Sun is still being debated, along with the methods used to determine this value, the general consensus is that active longitudes drift prograde relative to Carrington longitudes \citep[e.g.][]{ben99,uso07,ply10}.  To identify the drift rate of our longitudinal flux emergence patterns, we cross-correlate the emergence histograms of consecutive rotation periods for both the Northern and Southern hemispheres.  We average these cross-correlations for each hemisphere separately in order to identify an overall trend, as shown below in equation form for the Northern hemisphere:

\begin{equation}
CC_{Nh}=\frac{1}{18}\sum_{i=1}^{18} (Nh_{i} \star Nh_{i+1})
\end{equation} 

where the symbol $\star$ represents the cross-correlation, and $Nh_{i}$ ($Sh_{i}$) represents the i$^{th}$ emergence histogram for the Northern (Southern) hemisphere in a particular rotating reference frame.  Next we apply a 3-point running average to smooth the curve, then fit the curve with a Gaussian function to identify where a significant lead/lag occurs.  Such cross-correlations are shown in Figure \ref{fig:cc_cons} using three different rotating reference frames.  Values for the centers of the Gaussian fits in various reference frames are shown in Table \ref{tbl:table1} for both the Northern and Southern hemispheres in columns 2 and 3, respectively.   

In the reference frame rotating with angular velocity $\Omega_{0}/2\pi$, the Gaussian fit to the cross-correlation curves produce centers at small negative spatial lags.  These results indicate that the longitudinal flux emergence pattern in the $\Omega_{0}/2\pi$ reference frame propagate prograde at an average rate per rotation period of 4.2$^{\circ}$ in the Northern hemisphere, and 2.9$^{\circ}$ in the Southern hemisphere.  As this reference frame rotates at $\Omega_{0}/2\pi$, it is most likely the case that differential rotation present in the convection simulation helps to move the rising flux tube legs slightly prograde.  The prograde motion of the individual tubes contributes to a large-scale prograde drift of the flux emergence pattern between consecutive rotation periods.  In Paper 2 we find that with the addition of a time-varying convective velocity field, most flux tubes that emerge within $\pm15^{\circ}$ of the Equator rotate faster than the angular velocity $\Omega_{0}/2\pi$ of the simulation.  Although the large-scale flux emergence pattern in the reference frame rotating at $\Omega_{0}/2\pi$ does propagate prograde in each hemisphere, the pattern still rotates slower than the differentially rotating fluid in our convection simulation at a radius of 0.97R$_{\odot}$ (simulation upper boundary),  between latitudes of $\pm15^{\circ}$.         

Since the propagation rate of flux emergence patterns in the reference frame rotating at $\Omega_{0}/2\pi$ is only slightly prograde by a few degrees, it stands to reason that this flux emergence pattern will appear to propagate retrograde in a more rapidly rotating reference frame.  In the reference frame rotating at the ASH equivalent Carrington rate of $\Omega_{AC}/2\pi=461.70$ nHz, we find that the flux emergence pattern drifts retrograde by 13.7$^{\circ}$ per rotation period in the Northern hemisphere, and 12.0$^{\circ}$ in the Southern hemisphere between consecutive rotations.  Although this rotation is not prograde, as is most often the case for active longitudues on the Sun in the Carrington frame, these results show that the flux emergence patterns our simulation produces remain coherent for at least consecutive rotation periods in all reference frames.  Therefore, it is possible to identify a reference frame in which the flux emergence pattern appears to remain stationary.  We find the angular velocity of this reference frame to be $\Omega_{AL}/2\pi=440.64$ nHz, which corresponds to the ASH rotation rate at a depth of r=0.95R$_{\odot}$ and latitude $\theta=29^{\circ}$.  In finding this reference frame, we have also pinpointed the average drift rate of our flux emergence pattern.  It is important to note that the perceived prograde or retrograde drift rate of the flux emergence pattern depends on the details of the global differential rotation profile and the choice of reference frame, which are somewhat different in this ASH simulation relative to the Sun.

In the above analysis, we have included flux tubes of magnetic field strengths from 15 kG -100 kG.  However, taking the results of Paper 1 and Paper 2 into consideration, we suggest that the solar dynamo generated magnetic field strength is most likely between 40 kG - 100 kG.  In Figure \ref{fig:cc_cons_avg}, we show the average of all the cross-correlations between consecutive rotation periods in both the Northern and Southern hemispheres together, in three different reference frames for flux tubes of 15 kG - 100 kG (left), and 40 kG - 100 kG (right).  We have averaged the cross-correlations for the Northern and Southern hemispheres together ($CC_{Avg}=(CC_{Nh}+CC_{Sh})/2$)  in order to identify whether or not the drift rate of the flux emergence pattern changes for flux tubes of larger magnetic field strengths.  Since the difference between the centers of the Gaussian fits for the Northern and Southern hemispheres for each reference frame are not statistically significant, we feel it is valid to average the Northern and Southern hemisphere cross-correlation curves together for comparison between the magnetic field strength regimes.  

Values for the centers of the Gaussian fits to the average cross-correlation curves $CC_{Avg}$ are shown in columns 4 and 5 of Table \ref{tbl:table1}.  In the slowly rotating $\Omega_{0}/2\pi$ reference frame, the flux emergence pattern for 40 kG - 100 kG flux tubes drifts prograde in longitude an additional $6.1^{\circ}$ per rotation period compared to the flux emergence pattern of 15 kG - 100 kG flux tubes.  In the much faster ASH equivalent Carrington frame, $\Omega_{AC}/2\pi$, the flux emergence pattern for 40 kG - 100 kG flux tubes still rotates retrograde, but at a rate of only $10.6^{\circ}$ per rotation period compared to the $13.0^{\circ}$ for flux tubes of all magnetic field strengths.  This indicates that for 40 kG - 100 kG , the flux emergence pattern in the $\Omega_{AC}/2\pi$ frame moves forward in longitude faster than the 15 kG - 100 kG flux tube emergence pattern.  In Paper 2, we find that $\ge$60 kG flux tubes are capable of rotating at, or faster than, the inferred surface rate of our convection simulation at low latitudes of $\pm15^{\circ}$.  The faster prograde motion of 40 kG - 100 kG individual flux tubes contributed to a flux emergence pattern which rotates prograde in longitude compared to the 15 kG - 100 kG case.  In the $\Omega_{AL}/2\pi$ reference frame,  the flux emergence pattern still remains relatively stationary for 40 kG - 100 kG flux tubes.  However, Gaussian fits to the cross-correlation curves are rather broad, resulting in uncertainties which renders the difference in flux emergence drift rates between 15 kG - 100 kG flux tubes and 40 kG  - 100 kG flux tubes statistically insignificant. 

Our simulation produces a flux emergence pattern which rotates forward in longitude with respect to the reference frame rotating at $\Omega_{0}/2\pi=429.72$ nHz.  However, we do acknowledge that this pattern does not move prograde between consecutive rotation periods in the ASH equivalent Carrington frame $\Omega_{AC}/2\pi=461.72$ nHz, unlike active longitudes do on the Sun relative to the Carrington frame.  We attribute this to the fact that while the differential rotation of the convection simulation is very solar-like, it does not reproduce exactly the solar differential rotation profile.  Additionally, as found in Paper 2, simulated flux tubes of $\le$50 kG are not capable of rotating at, or faster than, the sunspot rotation rate.  The thin flux tube approximation breaks down near the top of the convection zone, so we cannot address the fact that the flux tube could lose its coherency and become fragmented in the upper convection zone.  It is possible that in these upper layers, the fragmented flux tube exhibits a stronger coupling to the convective fluid motions than is capable in this model, and could be a significant contributing factor to the rotation rate of flux tubes and their subsequent large-scale flux emergence pattern.

\subsection{Alignment of Flux Emergence Patterns Across the Equator}
\label{sec:align}

Occasionally an active zone is found on the Sun at the same longitude in both hemispheres \citep{ben99,detoma2000}.  Although the longitudes may be the same, the activity level of active longitudes in either hemisphere can vary greatly.  To identify the overall alignment trend of flux emergence in our simulation for a rotation period, we cross-correlate the emergence histograms of the Northern and Southern hemispheres for the same rotation period, then compute the average as follows:

\begin{equation}
CC_{NS}=\frac{1}{19}\sum_{i=1}^{19}(Nh_{i} \star Sh_{i})
\end{equation}  

In this way, common alignment trends will be amplified, and uncommon trends will be smeared out.  A  3-point running average is applied to smooth the cross-correlation result.  We then perform a Gaussian fit to the maximum, with the result of the fits for the three different reference frames in the 15 kG - 100 kG and 40 kG - 100 kG magnetic field strength regimes shown in Table \ref{tbl:table2}.  In all reference frames, the flux emergence patterns align very well across the Equator, regardless of magnetic field strength or rotation period.  These results are significant to the 99.7$\%$ confidence level.

All flux tubes in this simulation are perturbed with the exact same undular motions.  Although these perturbations are no longer needed because those provided by the convective velocity field are much stronger in amplitude, they have been included in order to facilitate comparison between flux tube properties with and without convection as presented in Paper 1 and Paper 2.  Removing the random phase relations from the thin flux tube model embedded in convection has a negligible effect on the dynamics and properties of the flux tube at all magnetic field strengths and magnetic flux.  Without convective effects, the flux tubes would emerge randomly distributed in longitude if perturbed initially with random undular motions.  Therefore, we would not expect to find a correlation between the North and South emergence histograms for the same rotation period.  

Even though we initiate identical flux tubes in the convection simulation at the same time in either hemisphere, the flux tubes do not evolve identically as each one is subjected to convective flows that are not symmetric across the Equator.  The fact that flux emergence patterns in our simulation align across the Equator suggests that the average near-hemispheric alignment trend of the longitudinal flux emergence pattern is convection dependent.  The different dominate modes in the power spectra found in Section \ref{sec:long_inhomo} for both hemispheres, $m=1$ in the Northern hemisphere and $m=3$ in the Southern hemisphere, do not contradict the findings in this Section.  On average, as long as the single active longitude in the Northern hemisphere aligns with any of the three in the Southern hemisphere, the correlation between hemispheric flux emergence patterns will be greater than what would be expected from random longitudinal flux tube emergence.  This effect is most probably related to the elongated downflow lanes present in our convection simulation, which span across the Equator and are rotationally aligned at low latitudes (see Fig. \ref{fig:mollweide}).


\section{Discussion}
\label{sec:discuss}

By coupling a thin flux tube model with a three-dimensional, solar-like, rotating convection simulation, we are able to examine the interaction of turbulent convective flows with buoyantly rising flux tubes.  As a result of this interaction, we find that flux tubes emerge at the upper boundary of our simulation domain in a large-scale longitudinal pattern which resembles the behavior of active longitudes on the Sun and other solar-like stars.  While the active longitude phenomenon most likely results from a combination of various physical mechanisms, we suggest that convection also has some part to play.

Flux emergence on the Sun and other solar-like stars is often concentrated in distinct bands with low order longitudinal modes.  The large-scale organization of convection in our simulation produces a flux emergence pattern that does exhibit low order longitudinal modes present at $m\le18$, consistently greater than the 99.7$\%$ confidence level, above which we are 99.7$\%$ sure the result is not due to a random distribution in longitude of flux tube emergence (Fig. \ref{fig:power}).  In the Northern hemisphere, the peak in the power spectrum occurs for longitudinal mode $m=1$, and $m=3$ in the Southern hemisphere.  However, the more important result is that both hemispheres exhibit secondary peaks in their power spectra for longitudinal modes of $m=6$.  This corresponds to the dominant longitudinal wavenumber of the convection simulation, which is representative of periodic, giant-cell convection.  The maxima in the power spectrum for either hemisphere are independent of reference frame and magnetic field strength.  This is also not the result of a non-axisymmetric ($m\ne0$) mode which a flux tube might develop due to the non-linear growth of the magnetic buoyancy instability, as we are only capable of recording the emergence position of the first apex of any flux tube that reaches the upper boundary of our simulation domain.  Rather, these longitudinal modes are imposed by the strong downflows at the boundaries of the giant cells in our convection simulation.  Additionally, the extent of the upflows determines the 'window' where flux tubes can emerge.  

\begin{figure}
\epsscale{.75}
\plotone{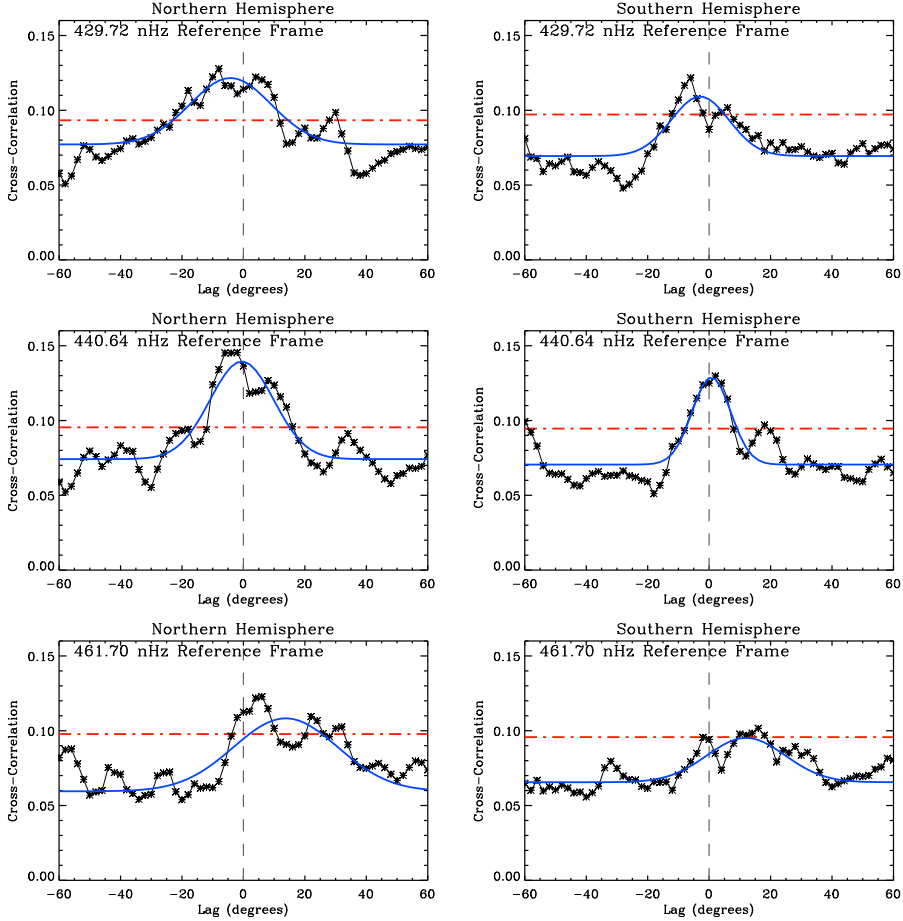}
\caption{Average cross-correlations of emergence histograms for the Northern ($CC_{Nh}$, left column) and Southern ($CC_{Sh}$, right column) hemispheres between consecutive rotation periods.  This has been done for three different rotating reference frames.  A Gaussian fit (blue line) to the maximum of the cross-correlation curves peak at or above the 99.7\% confidence level (3$\sigma$, red dash-dotted line), above which we are 99.7$\%$ certain the result is not due to a random distribution of flux tube emergence.  These curves indicate that the flux emergence pattern propagates at a steady rate and remains coherent between consecutive rotation periods.  Depending on the choice of rotating reference frame, the flux emergence pattern appears to drift prograde (top panels), retrograde (bottom panels), or remain stationary (middle panels).  The values of these drift rates are quoted in columns 2 and 3 of Table \ref{tbl:table1}.}
\label{fig:cc_cons}
\end{figure}

\begin{figure}
\epsscale{.75}
\plotone{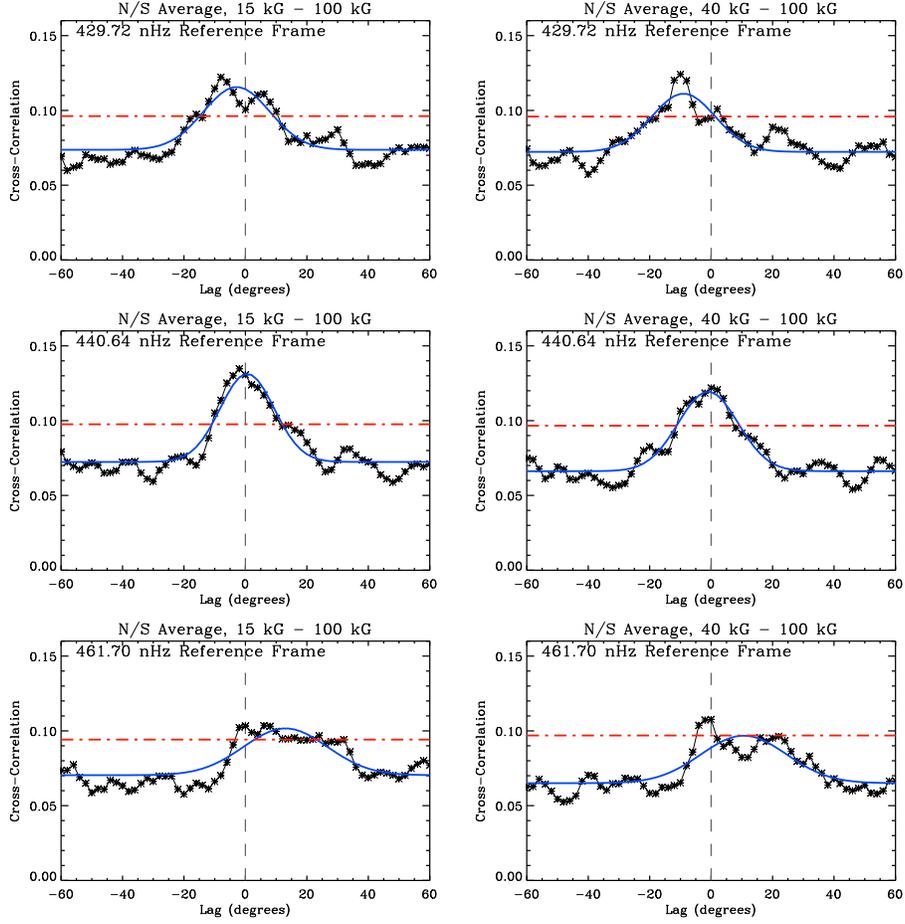}
\caption{Similar to Fig. \ref{fig:cc_cons}, but for the Northern and Southern hemispheres averaged together ($CC_{Avg}$), with the left column representing initial magnetic field strength of 15 kG - 100 kG, and the right column for 40 kG -100 kG flux tubes.  The values of these drift rates are quoted in columns 4 and 5 of Table \ref{tbl:table1}.  The center of the Gaussian fit is shifted prograde when flux tubes of 40 kG - 100 kG are considered.}
\label{fig:cc_cons_avg}
\end{figure}

Active longitudes on the Sun are known to drift prograde relative to a fixed reference system. The flux emergence patterns in our simulation propagate prograde relative to the mean angular velocity of our solar-like star, $\Omega_{0}/2\pi=429.72$ nHz.  Average drift rates are similar for each hemisphere, rotating at $4.2^{\circ}\pm3.6^{\circ}$ prograde in the Northern hemisphere, and $2.9^{\circ}\pm3.1^{\circ}$ in the Southern hemisphere (Fig. \ref{fig:cc_cons}, top) per rotation period.  However, in the more rapidly rotating ASH equivalent Carrington frame with $\Omega_{AC}/2\pi=461.72$ nHz, the flux emergence pattern propagates retrograde at $13.7^{\circ}\pm4.1^{\circ}$ in the Northern hemisphere, and $12.0^{\circ}\pm3.6^{\circ}$ in the Southern hemisphere (Fig. \ref{fig:cc_cons}, bottom).  We also identify a reference frame rotating at the angular velocity $\Omega_{AL}/2\pi=440.64$ nHz, in which the flux emergence pattern remains stationary, indicating that the flux emergence pattern in our simulation has an average angular velocity of $\sim$441 nHz.  This angular velocity corresponds to the rotation rate of the ASH simulation at r=0.95R$_{\odot}$ and a latitude of 29$^{\circ}$.  Although the flux emergence patterns in the $\Omega_{AC}/2\pi$ reference frame do not propagate prograde, they do have an average rotation rate of $\sim441$ nHz, which is faster than the mean angular velocity of the simulation, $\Omega_{0}/2\pi$.  Differential rotation present in the convection simulation helps to usher the flux emergence pattern forward in longitude.  The fact that we are capable of calculating cross-correlations that peak at or above the 99.7$\%$ confidence level means that the flux emergence pattern persists between at least consecutive rotation periods.  We also note that on average, the flux emergence pattern for any particular rotation period tends to align very closely across the Equator. This trend is related to the elongated downflow lanes which span across the Equator in the convection simulation (see Fig. \ref{fig:mollweide}).

\begin{figure}
\epsscale{.8}
\plotone{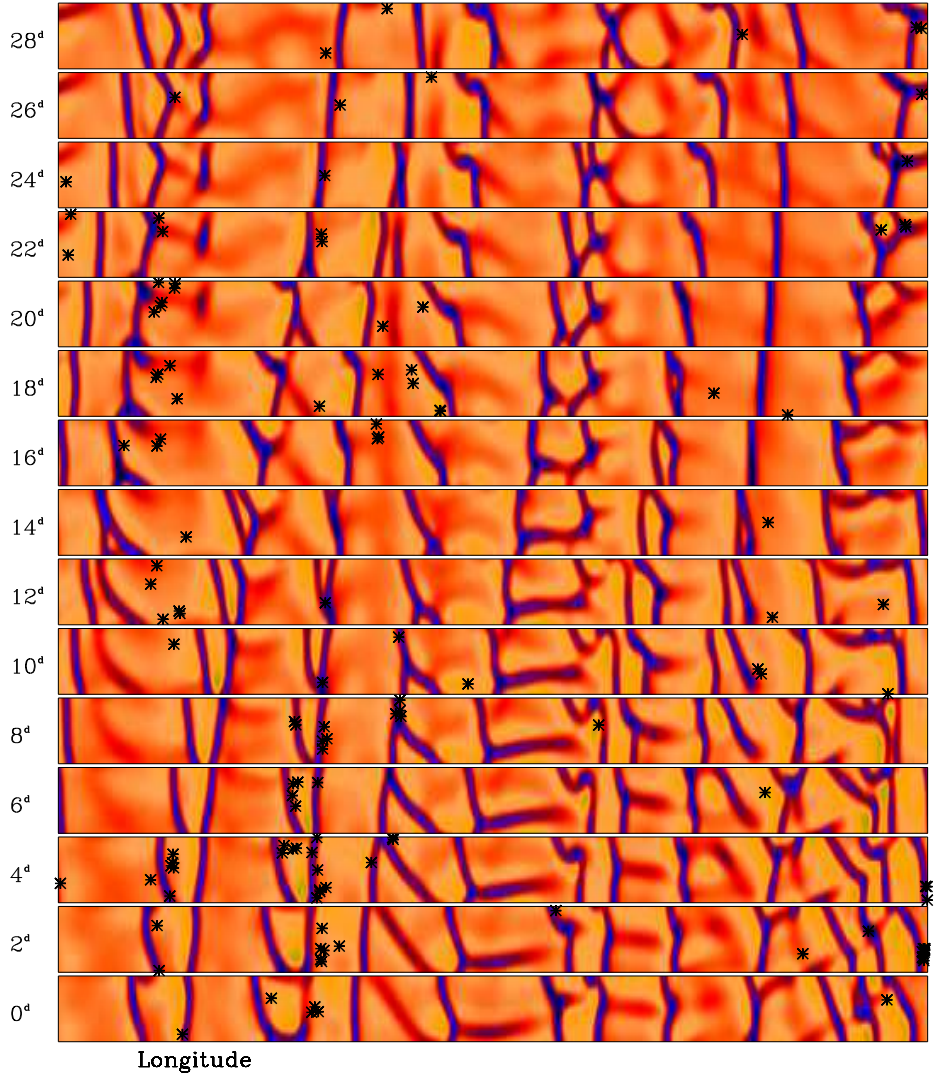}
\caption{ASH radial velocity snapshots at a depth of 0.95$R_{\odot}$ in strips of longitude (0$^{\circ}$ to 360$^{\circ}$) and latitude (0$^{\circ}$ to +15$^{\circ}$) for 28 consecutive days, at increments of every 2 days, in the reference frame rotating at the angular velocity $\Omega_{0}/2\pi$.  The initial starting time is arbitrary. Black asterisks mark the apex of all flux tubes which emerge within $\pm$1 day of the ASH snapshot.  This figure depicts how the nature of giant cells force flux tubes to emerge along the giant cell boundaries, and occasionally near the center.}
\label{fig:ashstack}
\end{figure}

Throughout this Paper, we highlight the ability of convection to organize flux emergence in a large-scale way.  The flux emergence pattern present in our simulations exhibits properties similar to those of active longitudes on the Sun.  We attribute the cause of these characteristics in our simulation to differential rotation and the columnar, rotationally aligned giant cells which exist in the equatorial to mid-latitude regions of our ASH convection simulation.  Convective downflows which mark the boundary of giant cells are capable of deforming even the strongest flux tubes, forcing them to emerge along the boundaries of the giant cells, and occasionally near the center as depicted in Figure \ref{fig:ashstack}.  These flux tubes tend to emerge at r=0.97R$_{\odot}$ near the giant cell boundaries because the strongest upflows occur near the downflow lanes \citep{miesch08}.  Additionally, there is a positive horizontal divergence of the velocity field within the upflow region of the giant cell, which helps expel the rising flux tube toward the cell boundary \citep{miesch08}.  

Rising flux tubes are subject not only to the mean flows of the convection simulation, such as differential rotation, but also to the prograde propagation of the ever changing giant cells near the Equator.  Although continually evolving, some of the giant cells are capable of remaining coherent for a rotation period or longer (see Fig. \ref{fig:ash_strip}).   The giant cells 'corral' flux tubes in a sense, forcing them to emerge within the boundaries of a particular cellular feature.  The periodic nature of the giant cells in longitude is most likely the reason why the flux emergence patterns in our simulation exhibit low order longitudinal modes, as the downflow lanes can restrict the rise of flux tubes over certain longitudinal spans.  Differential rotation aids these distinct longitudinal bands of flux in rotatating prograde in longitude.  Downflow lanes of these giant cells extend across the Equator and are rotationally aligned, forcing the alignment of the flux emergence pattern across the Equator, which is depicted well in the flux tube emergence maps for both hemispheres in Figure \ref{fig:emergence_map}.  The sum of these results suggest to us that giant cells may play a significant role in the active longitude phenomena on the Sun and other solar-like stars.

\section{Acknowledgements}
This work is supported in part by NASA SHP grant NNX10AB81G to the National Center for Atmospheric Research (NCAR).  NCAR is sponsored by the National Science Foundation.  We would also like to thank Matthias Rempel, Giuliana de Toma, Masumi Dikpati, and Nick Featherstone for their suggestions, help, and advice.



\begin{table}
\begin{tabular}{cccccc}
\hline
\hline
 Angular Velocity  & North & South & Avg N/S & Avg N/S \\
$\Omega/2\pi$ (nHz) & 15 kG - 100 kG & 15 kG - 100 kG & 15 kG - 100 kG  & 40 kG - 100 kG\\

\hline
$\Omega_{0}/2\pi$, 429.72   	&		$-4.2^{\circ}\pm3.6^{\circ}$    
						&		$-2.9^{\circ}\pm3.1^{\circ}$ 
 		&		$-2.9^{\circ}\pm3.3^{\circ}$    &		$-9.0^{\circ}\pm3.2^{\circ}$   \\
$\Omega_{AL}/2\pi$, 440.64   	& 		$-0.5^{\circ}\pm3.2^{\circ}$    
						&		$0.7^{\circ}\pm2.4^{\circ}$   
             	&		$0.6^{\circ}\pm3.0^{\circ}$      &		$-1.0^{\circ}\pm3.1^{\circ}$   \\
$\Omega_{AC}/2\pi$, 461.70   	&		$13.7^{\circ}\pm4.1^{\circ}$   
						&		$12.0^{\circ}\pm3.6^{\circ}$  
                 	&		$13.0^{\circ}\pm3.7^{\circ}$   &		$10.6^{\circ}\pm3.7^{\circ}$  \\
\hline

\end{tabular}
\caption{Centers of the Gaussian fit to the average of the cross-correlations for 19 consecutive rotation periods are shown in columns 2 ($CC_{Nh}$) and 3 ($CC_{Sh}$).  In columns 4 and 5, the cross-correlations for the Northern and Southern hemispheres are averaged together ($CC_{Avg}$) for different magnetic field strength regimes.  Uncertainties are the standard deviation of the Gaussian function.  This is shown for three different reference frames rotating at angular velocities given in column 1.  Negative values imply prograde propagation, and positive values imply retrograde propagation.}
\label{tbl:table1}
\end{table}

\begin{table}
\begin{tabular}{cccc}
\hline
\hline
$\Omega/2\pi$ (nHz) & 15 kG - 100 kG & 40 kG - 100 kG\\

\hline
$\Omega_{0}/2\pi$, 429.72 	&	$1.1^{\circ}\pm2.7^{\circ}$		
						&	$-1.5^{\circ}\pm2.1^{\circ}$  \\
$\Omega_{AL}/2\pi$, 440.64    	& 	$1.9^{\circ}\pm2.9^{\circ}$    	
						&	$-4.1^{\circ}\pm2.6^{\circ}$    \\
$\Omega_{AC}/2\pi$, 461.70    	&	$0.5^{\circ}\pm2.3^{\circ}$   	
						&	$0.1^{\circ}\pm2.0^{\circ}$  \\
\hline

\end{tabular}
\caption{Centers of the Gaussian fit for cross-correlations of emergence histograms of the average of the Northern and Southern hemispheres for the same rotation period ($CC_{NS}$).  Uncertainties are the standard deviation of the Gaussian function.  These are shown for three different reference frames rotating at angular velocities given in column 1.}
\label{tbl:table2}
\end{table}

\end{document}